\documentclass[11pt,a4paper]{article}
\usepackage{acl2015}
\usepackage[ruled,norelsize]{algorithm2e}
\SetAlFnt{\small}
\SetAlCapFnt{\small}
\SetAlCapNameFnt{\small}
\usepackage[fleqn]{amsmath}
\usepackage{algorithmicx}[1]
\usepackage{amssymb}
\usepackage{times}
\usepackage{url}
\usepackage{latexsym}
\usepackage{colortbl}
\usepackage{multirow}
\usepackage{xcolor}
\usepackage{booktabs}
\usepackage[fleqn]{mathtools}
\usepackage{paralist}
\usepackage[utf8]{inputenc}
\usepackage{enumitem}
\usepackage{xspace}
\usepackage[font=normal,format=plain,labelfont=up,textfont=up,justification=justified,singlelinecheck=false]{caption}

\DeclarePairedDelimiterX{\infdivx}[2]{(}{)}{%
  #1\;\delimsize\|\;#2%
}
\newcommand{\infdiv}{KL\infdivx}
\DeclarePairedDelimiter{\norm}{\lVert}{\rVert}

\DeclareMathOperator*{\argmin}{arg\,min}

\title{Semantic Annotation for Microblog Topics\\Using Wikipedia Temporal Information}

\author{
  Tuan Tran \\
  L3S Research Center \\
  Hannover, Germany\\
  {\tt ttran@L3S.de} \\
  \And
  Nam Khanh Tran \\
  L3S Research Center \\
  Hannover, Germany\\
  {\tt ntran@L3S.de} \\
  \And
  Asmelash Teka Hadgu \\
  L3S Research Center \\
  Hannover, Germany\\
  {\tt teka@L3S.de} \\
  \And
  Robert Jäschke \\
  L3S Research Center \\
  Hannover, Germany\\
  {\tt jaeschke@L3S.de} \\
}

\date{}

\definecolor{atk}{rgb}{0.698,0.122,0.435}
\definecolor{rja}{rgb}{0.122,0.435,0.698}
\definecolor{tuan}{rgb}{0.435,0.698,0.122}
\definecolor{nam}{rgb}{0.300,0.227,0.050}
\definecolor{TODO}{rgb}{0.784,0.145,0.00}

\newcommand{\final}[1]{\textbf{/* for camera ready/long version: #1  */}}

\renewcommand{\final}[1]{}

\newcommand{\parai}[1]{\paragraph{#1}}

\newcommand{\wpent}[1]{\emph{#1}\xspace}

\begin{document}
\maketitle
\begin{abstract}
  Trending topics in microblogs such as Twitter are valuable resources
  to understand social aspects of real-world events. To enable deep
  analyses of such trends, semantic annotation is an effective
  approach; yet the problem of annotating microblog trending topics is
  largely unexplored by the research community. In this work, we
  tackle the problem of mapping trending Twitter topics to entities
  from Wikipedia. We propose a novel model that complements
  traditional text-based approaches by rewarding entities that exhibit
  a high temporal correlation with topics during their burst time
  period. By exploiting temporal information from the Wikipedia edit
  history and page view logs, we have improved the annotation
  performance by 17-28\%, as compared to the competitive baselines.
\end{abstract}

\section{Introduction}\label{sec:intro}


With the proliferation of microblogging and its wide influence on how
information is shared and digested, the studying of microblog sites
has gained interest in recent NLP research. Several approaches have
been proposed to enable a deep understanding of information on
Twitter. An emerging approach is to use semantic annotation
techniques, for instance by mapping Twitter information snippets to
canonical entities in a knowledge base or to
Wikipedia~\cite{meij2012adding,guo2013link}, or by revisiting NLP
tasks in the Twitter
domain~\cite{owoputi2013improved,ritter2011named}. Much of the
existing work focuses on annotating a single Twitter message
(tweet). However, information in Twitter is rarely digested in
isolation, but rather in a collective manner, with the adoption of
special mechanisms such as hashtags. When put together, the
unprecedentedly massive adoption of a hashtag within
a short time period can lead to bursts and often reflect trending
social attention.
Understanding the meaning of trending hashtags offers a valuable
opportunity for various applications and studies, such as viral
marketing, social behavior analysis, recommendation,
etc. Unfortunately, the task of hashtag annotation has been largely
unexplored so far.

\begin{figure}[!tb]
  \centering
  \includegraphics[width=\columnwidth]{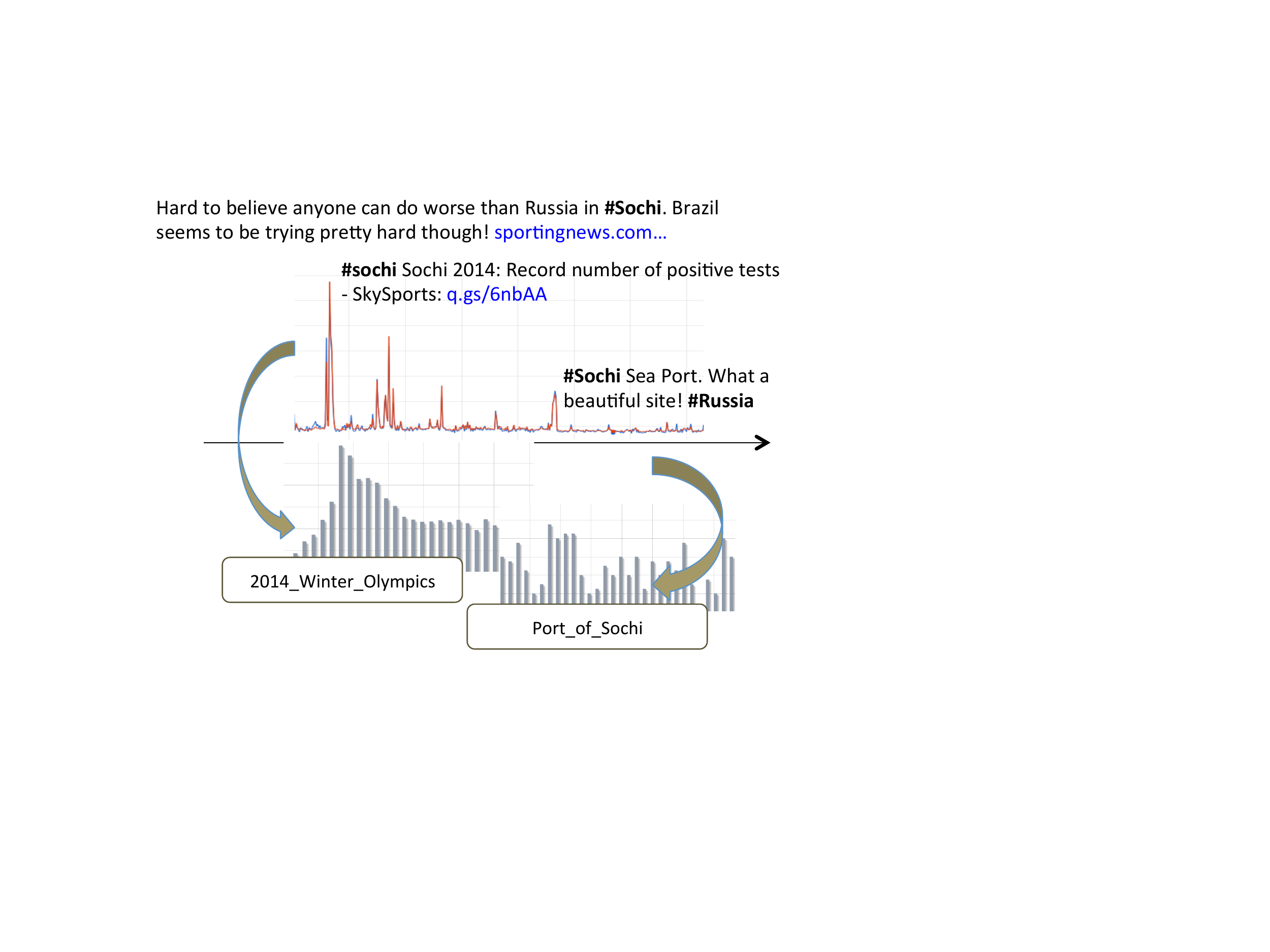}
  \caption{Example of trending hashtag annotation. During the
    \wpent{2014 Winter Olympics}, the hashtag `\#sochi' had a
    different meaning.}\label{fig:sochi}
\end{figure}

In this paper, we study the problem of annotating trending hashtags on
Twitter by entities derived from Wikipedia.  Instead of establishing a
static semantic connection between hashtags and entities, we are
interested in \emph{dynamically} linking the hashtags to entities that
are closest to the underlying topics during burst time periods of
the hashtags.
For instance, while `\#sochi' refers to a city in Russia, during
February 2014, the hashtag was used to report the \wpent{2014 Winter
  Olympics} (cf. Figure~\ref{fig:sochi}). Hence, it should be
linked more to Wikipedia pages related to the event than to the
location.

Compared to traditional domains of text (e.g., news articles),
annotating hashtags poses additional challenges. Hashtags'
surface forms are very ad-hoc, as they are chosen not in favor of the
text quality, but by the dynamics in attention of the large crowd. In
addition, the evolution of the semantics of hashtags (e.g., in
the case of `\#sochi') makes them more ambiguous. 
Furthermore, a hashtag can encode multiple topics at once.
For example, in March 2014, `\#oscar' refers to the \wpent{86th
  Academy Awards}, but at the same time also to the \wpent{Trial of
  Oscar Pistorius}.
Sometimes, it is difficult even for humans to understand a trending
hashtag without knowledge about what was happening with the related entities in
the real world.

In this work, we propose a novel solution to these challenges by
leveraging temporal knowledge about entity dynamics derived from
Wikipedia. We hypothesize that a trending hashtag
is associated with an increase in public attention to certain entities, and
this can also be observed on Wikipedia.
As in Figure~\ref{fig:sochi}, we can identify \wpent{2014 Winter
  Olympics} as a prominent entity for `\#sochi' during February 2014,
by observing the change of user attention to the entity, for instance
via the page view statistics of Wikipedia articles.
We exploit both Wikipedia edits and page views for annotation. We also
propose a novel learning method, inspired by the information spreading
nature of social media such as Twitter, to suggest the optimal
annotations without the need for human labeling. In summary:
\begin{itemize}
\item We are the first to combine the Wikipedia edit history and page
  view statistics to overcome the temporal ambiguity of Twitter
  hashtags.
\item We propose a novel and efficient learning algorithm based on
  influence maximization to automatically annotate hashtags. The idea
  is generalizable to other social media sites that have a similar
  information spreading nature.
\item We conduct thorough experiments on a real-world dataset and show
  that our system can outperform competitive baselines by 17-28\%.
\end{itemize}


\section{Related Work}

\paragraph{Entity Linking in Microblogs} The task of semantic
annotation in microblogs has been recently tackled by different
methods,
which
can be divided into two classes, i.e., content-based and graph-based
methods.  While the content-based methods
\cite{meij2012adding,guo2013link,fang2014entity}
consider tweets independently, the graph-based methods
\cite{cassidy2012analysis,liu2013entity} use all related tweets (e.g.,
posted by a user) together.  However, most of them focus on entity
mentions in tweets. In contrast, we take into account hashtags which
reflect the topics discussed in tweets, and leverage external
resources from Wikipedia (in particular, the edit history and page view
logs) for semantic annotation.

\paragraph{Analysis of Twitter Hashtags} In an attempt to understand
the user interest dynamics on Twitter, a rich body of work analyzes
the temporal patterns of popular hashtags
\cite{lehmann2012dynamical,naaman2011hip,Tsur:2012}.
Few works have paid attention to the semantics of hashtags, i.e., to
the underlying topics conveyed in the corresponding tweets. Recently,
\newcite{BansalBV15} attempt to segment a hashtag and link each of its tokens
to a Wikipedia page. However, the authors only aim to retrieve
entities directly mentioned within a hashtag, which are very few in
practice. The external information derived from the tweets is largely
ignored. In contrast, we
exploit both context
information from the microblog and Wikipedia resources.

\paragraph{Event Mining Using Wikipedia} 
Recently some works exploit Wikipedia for detecting and analyzing
events on Twitter
\cite{osborne2012bieber,Tolomei2013TAB,Tran2014Wikitweet}.  However,
most of the existing studies focus on the statistical signals of
Wikipedia (such as the edit or page view volumes). We are the first to
combine the content of the Wikipedia edit history and the magnitude of
page views to handle trending topics on Twitter.


\section{Framework}\label{sec:framework}

\paragraph{Preliminaries} We refer to an \emph{entity} (denoted
by~$e$) as any object described by a Wikipedia article (ignoring
disambiguation, lists, and redirect pages). The number of times an
entity's article has been requested is called the \emph{entity view
  count}. The text content of the article is denoted by $C(e)$.
In this work,
we choose to study hashtags at the daily level, i.e., from the
timestamps of tweets we only consider their creation day. 
A hashtag is called \emph{trending} at a time point (a day) if the number of
tweets where it appears is significantly higher than that on other days.  There are many ways to detect such trendings.
\cite{Lappas:2009,lehmann2012dynamical}. Each trending hashtag has one
or multiple \emph{burst time periods}, surrounding the trending day, where the users' interest in the underlying topic remains
stronger than in other periods. We denote with $T(h)$ (or $T$ for
short) one hashtag burst time period,
and with $D_T(h)$ the set of tweets containing the hashtag $h$ created
during $T$.
%
%
%

\paragraph{Task Definition} Given a trending hashtag $h$ and the burst
time period $T$ of $h$, identify the top-$k$ most prominent entities
to describe $h$ during $T$.

It is worth noting that not all trending hashtags are mapable to
Wikipedia entities, as the coverage of topics in Wikipedia is much
lower than on Twitter. This is also the limitation of systems relying on
Wikipedia such as entity disambiguation, which can only disambiguate
popular entities and not the ones in the long tail. In this study, we
focus on the precision and the popular trending hashtags, and leave
the improvement of recall to future work.

\paragraph{Overview} We approach the task in three steps. The first
step is to identify all entity candidates by checking surface forms of
the constituent tweets of the hashtag. In the second step, we compute
different similarities between each candidate and the hashtag, based
on different types of contexts, which are derived from either side
(Wikipedia or Twitter). Finally, we learn a unified ranking function
for each (hashtag, entity) pair and choose the top-$k$ entities with
the highest scores. The ranking function is learned through an
unsupervised model and needs no human-defined labels.

\subsection{Entity Linking}\label{subsec:el}
The most obvious resource to identify candidate entities for a hashtag
is via its tweets. We follow common approaches that use a lexicon to
match each textual phrase in a tweet to a potential entity set
\cite{shen2013linking,fang2014entity}. Our lexicon is constructed from
Wikipedia page titles, hyperlink anchors, redirects, and
disambiguation pages, which are mapped to the corresponding entities.
As for the tweet phrases, we extract all $n$-grams ($n\leq 5$) from
the input tweets within $T$. We apply the longest-match heuristic
\cite{meij2012adding}: We start with the longest $n$-grams and stop as
soon as the entity set is found, otherwise we continue with the smaller
constituent $n$-grams.

\parai{Candidate Set Expansion} While the lexicon-based linking works well
for single tweets, applying it on the hashtag level has subtle
implications.
Processing a huge amount of text, especially during a hashtag burst
time period, incurs expensive computational costs. Therefore, to
guarantee a good recall in this step while still maintaining
feasible computation, we apply entity linking only on a random sample
of the complete tweet set.
Then, for each candidate entity $e$, we include all entities whose
Wikipedia article is linked with the article of $e$ by an outgoing or
incoming link.

\subsection{Measuring Entity--Hashtag Similarities}
To rank the entity by prominence, we measure the similarity between
each candidate entity and the hashtag. We study three types of
similarities:
%

\parai{Mention Similarity} This measure relies on the explicit
mentions of entities in tweets. It assumes that entities directly
linked from more prominent anchors are more relevant to the
hashtag. It is estimated using both statistics from Wikipedia and
tweet phrases, and turns out to be surprisingly effective in practice~\cite{fang2014entity}.

\parai{Context Similarity}
For entities that are not directly linked to mentions (the mention
similarity is zero) we exploit external resources instead. Their
prominence is perceived by users via external sources, such as web
pages linked from tweets, or entity home pages or Wikipedia pages. By exploiting the
content of entities from these external sources, we can complement the
explicit similarity metrics based on mentions.

\parai{Temporal Similarity}
The two measures above rely on the textual representation
and are degraded
by the linguistic difference between the two platforms. To overcome
this drawback, we incorporate the temporal dynamics of hashtags and
entities, which serve as a proxy to the change of user interests
towards the underlying topics~\cite{ciglan2010wikipop}. We employ the
correlation between the times series of hashtag adoption and the
entity view as the third similarity measure.

\subsection{Ranking Entity Prominence}
While each similarity measure captures one evidence of the entity prominence, we need to unify all scores to obtain a global ranking function. In this work, we propose to combine the individual similarities using a linear function: 
\begin{equation}\label{eq:of}
  f(e,h) = \alpha f_m(e,h)+\beta f_c(e,h)+\gamma f_t(e,h)
\end{equation}
where $\alpha,\beta,\gamma$ are model weights and $f_m,f_c,f_t$ are
the similarity measures based on mentions, context, and temporal
information, respectively, between the entity $e$ and the hashtag
$h$. We further constrain that $\alpha+\beta+\gamma=1$, so that the
ranking scores of entities are normalized between $0$ and $1$, and
that our learning algorithm is more tractable.  The algorithm, which
automatically learns the parameters without the need of human-labeled
data, is explained in detail in Section~\ref{sec:im}.


\begin{figure*}[ht]
  \centering
  \includegraphics[width=2\columnwidth]{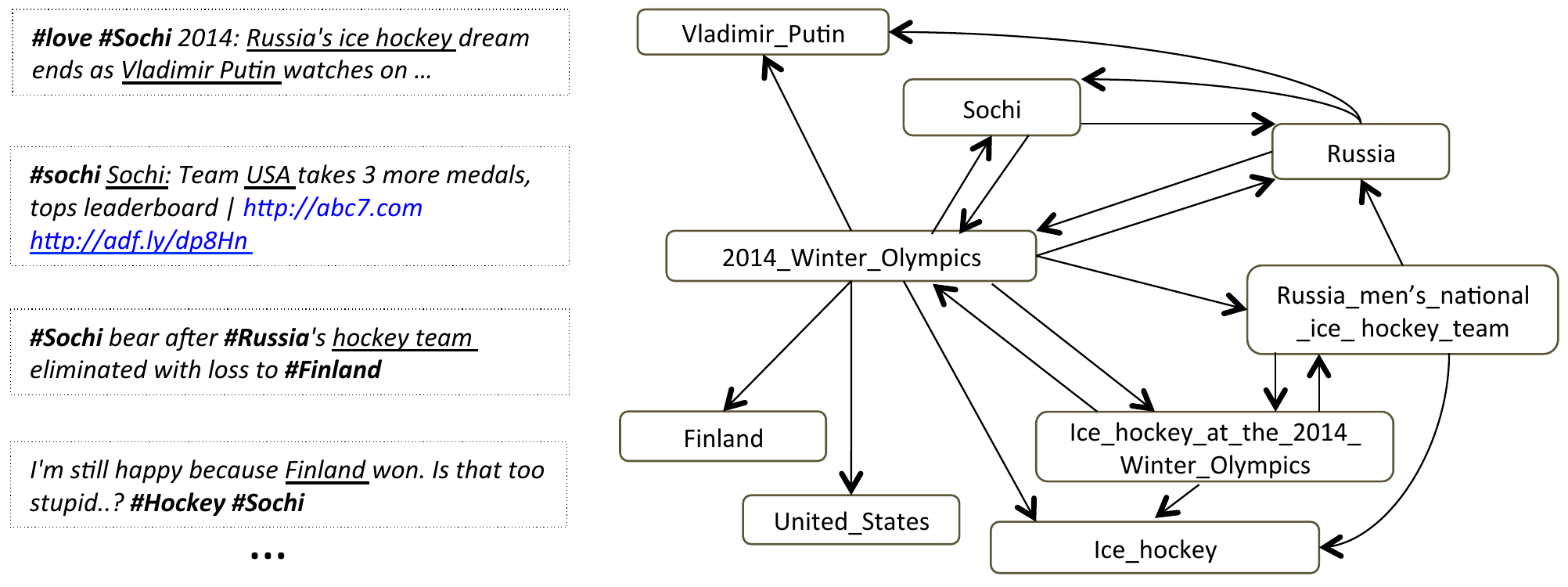}
  \caption{Excerpt of tweets about ice hockey results in the
    \wpent{2014 Winter Olympics} (left), and the observed linking
    process between time-aligned revisions of candidate Wikipedia
    entities (right). Links come more from prominent entities to
    marginal ones to provide background, or more context for the
    topics. Thus, starting from prominent entities, we can reach more
    entities in the graph of candidate entities}\label{fig:infgraph}
\end{figure*}

\section{Similarity Measures}\label{sec:similarity-measures}

We now discuss in detail how the similarity measures between hashtags
and entities are computed.

\subsection{Link-based Mention Similarity}
The similarity of an entity with one individual mention in a tweet can
be interpreted as the probabilistic prior in mapping the mention to
the entity via the lexicon. One common way to estimate the entity
prior exploits the anchor statistics from Wikipedia links, and has
been proven to work well in different domains of text. We follow this
approach and define $LP(e|m)=\frac{|l_m(e)|}{\sum_{m'}|l_{m'}(e)|}$ as
the link prior of the entity $e$ given a mention $m$, where $l_m(e)$
is the set of links with anchor $m$ that point to $e$. The mention
similarity $f_m$ is measured as the aggregation of link priors of the
entity $e$ over all mentions in all tweets with the hashtag $h$:
\begin{equation}\label{eq:m}
  f_m(e,h)=\sum_{m}(LP(e|m)\cdot q(m))
\end{equation}
where $q(m)$ is the frequency of the mention $m$ over all mentions of $e$ in all tweets of
$h$.

\subsubsection{Context Similarity}\label{subsec:cs}
To compute $f_c$, we first construct the contexts for hashtags and
entities. The context of a hashtag is built by extracting all words
from its tweets. We tokenize and parse the tweets' part-of-speech
tags \cite{owoputi2013improved}, and remove words
of Twitter-specific tags (e.g., @-mentions, URLs, emoticons,
etc.). Hashtags are normalized using the word breaking method by
\newcite{wang2011web}. 

The textual context of an entity is
extracted from its Wikipedia article. One subtle aspect is that the
articles are not created at once, but are incrementally updated over
time in accordance with changing information about entities. Texts
added in the same time period of a trending hashtag contribute more to
the context similarity between the entity and the hashtag.
Based on this observation, we use the Wikipedia revision history -- an
archive of all revisions of Wikipedia articles -- to calculate the
entity context. We collect the revisions of articles during the time
period $T$, plus one day to acknowledge possible time lags. We compute
the difference between two consecutive revisions, and extract only the
added text snippets. These snippets are accumulated to form the
\emph{temporal context} of an entity $e$ during $T$, denoted by
$C_T(e)$. The distribution of a word $w$ for the entity $e$ is
estimated by a mixture between the probability of generating $w$ from
the temporal context and from the general context $C(e)$ of the
entity:
\vspace{-0.3em}
\begin{equation*}
  \hat{P}(w|e) = \lambda \hat{P}(w|M_{C_T(e)}) + (1-\lambda) \hat{P}(w|M_{C(e)})
\end{equation*}
where $M_{C_T(e)}$ and $M_{C(e)}$ are the language models of $e$ based
on $C_T(e)$ and $C(e)$,
respectively. The probability $\hat{P}(w|M_{C(e)})$ can be regarded as
corresponding to the background model, while $\hat{P}(w|M_{C_T(e)})$
corresponds to the foreground model in traditional language modeling
settings. Here we use a simple maximum likelihood estimation to
estimate these probabilities:
$\hat{P}(w|M_{C(e)})=\frac{tf_{w,c}}{|C(e)|}$
and
$\hat{P}(w|M_{C_T(e)})=\frac{tf_{w,c_T}}{|C_T(e)|}$, 
where $tf_{w,c}$ and $tf_{w,c_T}$ are the term frequencies of $w$ in
the two text sources of $C(e)$ and $C_T(e)$, respectively, and
$|C(e)|$ and $|C_T(e)|$ are the lengths of the two texts,
respectively. We use the same estimation for tweets: 
$\hat{P}(w|h) = \frac{tf_{w,D(h)}}{|D(h)|}$,
where $D(h)$ is the concatenated text of all tweets of $h$ in $T$.
We use and normalize the Kullback-Leibler divergence to compare
the distributions over all words appearing both in the Wikipedia
contexts and the tweets:
\begin{align}\label{eq:c}
  \infdiv{e}{h} &= \sum_w \hat{P}(w|e)\cdot\frac{\hat{P}(w|e)}{\hat{P}(w|h)} \nonumber \\
  f_c(e,h) &= e^{-\infdiv{e}{h}} &
\end{align}

\subsubsection{Temporal Similarity}
The third similarity, $f_t$, is computed using temporal signals from
both sources -- Twitter and Wikipedia. For the hashtags, we build the
time series based on the volume of tweets adopting the hashtag $h$ on
each day in $T$: $TS_h = [n_1,n_2,\ldots,n_{|T|}]$. Similarly for the
entities,
we build the time series of view counts for the entity $e$ in $T$:
$TS_e = [v_1,v_2,\ldots,v_{|T|}]$. A time series similarity metric is
then used to compute $f_t$. Several metrics can be used, however most
of them suffer from the time lag and scaling discrepancy, or incur
expensive computational costs \cite{Radinsky2011}. In this work, we
employ a simple yet effective metric that is agnostic to the scaling
and time lag of time series \cite{yang2011patterns}. It measures the
distance between two time series by finding optimal shifting and
scaling parameters to match the shape of two time series:
\begin{equation}\label{eq:t}
  f_t(e,h) = \min_{q,\delta}\frac{\norm{TS_h - \delta d_q(TS_e)}}{\norm{TS_h}}
\end{equation}
where $d_q(TS_e)$ is the time series derived from $TS_e$ by shifting
$q$ time units, and $\norm{\cdot}$ is the $L_2$ norm. It has been
proven that Equation~\ref{eq:t} has a closed-form solution for
$\delta$ given fixed $q$, thus we can design an efficient
gradient-based optimization algorithm to compute $f_t$
\cite{yang2011patterns}.


\section{Entity Prominence Ranking}\label{sec:im}

\subsection{Ranking Framework}
To unify the individual similarities into one global metric
(Equation~\ref{eq:of}), we need a guiding premise of what manifest the
prominence of an entity to a hashtag.
Such a premise can be instructed through manual
assessment~\cite{meij2012adding,guo2013link}, but it requires
human-labeled data and is biased from evaluator to evaluator. Other
heuristics assume that
entities close to the main topic of a text are also coherent to each
other \cite{Ratinov:2011:LGA:2002472,liu2013entity}.  Based on this,
state-of-the-art methods in traditional disambiguation estimate entity prominence by optimizing the overall coherence of the
entities' semantic relatedness.
However, this coherence does not hold for topics in hashtags: Entities
reported in a big topic such as the Olympics vary greatly with
different sub-events. They are not always coherent to each other, as
they are largely dependent on the users' diverse attention to each
sub-event.
This heterogeneity of hashtags calls for a different premise,
abandoning the idea of coherence.

\paragraph{Influence Maximization (IM)} We propose a new approach to
find entities for a hashtag. We use an observed behavioral pattern in
creating Wikipedia pages for guiding our approach to entity
prominence: Wikipedia articles of entities that are prominent for a
topic are quickly created or
updated,\footnote{\newcite{osborne2012bieber} suggested a time lag of
  3~hours.} and subsequently enriched with links to related
entities. This linking process signals the dynamics of editor
attention and exposure to the
event~\cite{Keegan:2011:HOW:2038558.2038577}. We argue that the
process does not, or to a much lesser degree, happen to more marginal
entities or to very general entities. As illustrated in
Figure~\ref{fig:infgraph}, the entities closer to the 2014 Olympics
get more updates in the revisions of their Wikipedia articles, with
subsequent links pointing to articles of more distant entities. The
direction of the links influences the shifting attention of
users~\cite{Keegan:2011:HOW:2038558.2038577}
as they follow the structure of articles in Wikipedia.

We assume that, similar to Wikipedia, the entity prominence also
influences how users are exposed and spread the hashtag on Twitter. In
particular, the initial spreading of a trending hashtag involves more
entities in the focus of the topic. Subsequent exposure and spreading
of the hashtag then include other related entities (e.g., discussing
background or providing context), driven by interests in different
parts of the topic.
Based on this assumption, we propose to gauge the entity prominence
as its potential in~\emph{maximizing the information spreading} within
all entities present in the tweets of the hashtag. In other words, the
problem of ranking the most prominent entities becomes identifying the
set of entities that lead to the largest number of entities in the
candidate set. This problem is known in social network research as
\emph{influence maximization}~\cite{kempe2003maximizing}.


\paragraph{Iterative Influence-Prominence Learning (IPL)} IM itself is
an NP-hard problem~\cite{kempe2003maximizing}. Therefore, we propose an
approximation framework, which can \emph{jointly} learn the influence scores
of the entity and the entity prominence together. The framework
(called IPL) contains several iterations, each consisting of two
steps:%
\begin{inparaenum}[(1)]
\item Pick up a model and use it to compute the
  entity influence score.
\item Based on the influence scores, update the entity prominence.
\end{inparaenum}
In the sequel we detail our learning framework.

\subsection{Entity Graph}\label{subsec:la}

\paragraph{Influence Graph} To compute the entity influence scores, we
first construct the entity \emph{influence graph} as follows. For each
hashtag $h$, we construct a directed graph $G_h=(E_h,V_h)$, where the
nodes $E_h\subseteq E$ consist of all candidate entities
(cf. Section~\ref{subsec:el}), and an edge $(e_i,e_j)\in V_h$
indicates that there is a link from $e_j$'s Wikipedia article to
$e_i$'s.  Note that edges of the influence graph are inversed in
direction to links in Wikipedia, as such a link gives an ``influence
endorsement'' from the destination entity to the source entity.

\paragraph{Entity Relatedness} In this work, we assume that an entity
endorses more of its influence score to highly related entities than to lower related
ones. We use a popular entity relatedness measure suggested by 
\newcite{milne2008learning}:
\begin{equation*}
  \resizebox{\hsize}{!}{$MW(e_1,e_2) = 1 - \frac{\log(\max(|I_1|,|I_2|)-\log(|I_1\cap I_2|)))}{\log(|E|)-\log(\min(|I_1|,|I_2|))}$}
\end{equation*}
%
where $I_1$ and $I_2$ are sets of entities having links to $e_1$ and
$e_2$, respectively, and $E$ is the set of all entities in
Wikipedia. The influence transition from $e_i$ to $e_j$ is defined as the normalized value:
\begin{equation}
  b_{i,j} = \frac{MW(e_i,e_j)}{\sum_{(e_i,e_k)\in V} MW(e_i,e_k)}
\end{equation}


\paragraph{Influence Score} Let $\bold{r}_h$ be the influence score
vector of entities in $G_h$. We can estimate $\bold{r}_h$ efficiently
using random walk models, similarly
to the baseline method suggested by~\newcite{liu2014influence}:
\begin{equation}\label{eq:rw}
  \bold{r_h} \coloneqq \tau\bold{B}\bold{r_h} + (1-\tau)\bold{s_h}
\end{equation}
where $\bold{B}$ is the influence transition matrix, $\bold{s_h}$ are
the initial influence scores that are based on the entity prominence
model (Step~1 of IPL), and $\tau$ is the damping factor.

\subsection{Learning Algorithm}
Now we detail the IPL algorithm. The objective is to learn the model
$\omega=(\alpha,\beta,\gamma)$ of the global function
(Equation~\ref{eq:of}). The general idea is that we find an optimal
$\omega$ such that the average error with respect to the top
influencing entities is minimized
\begin{flalign*}
  \omega &= \argmin
  \sum_{E(h,k)}{L(f(e,h),r(e,h))}
\end{flalign*}
where $r(e,h)$ is the influence score of $e$ and $h$, $E(h,k)$ is the
set of top-$k$ entities with highest $r(e,h)$, and $L$ is the squared
error loss function, $L(x,y)=\frac{(x-y)^2}{2}$.

\begin{algorithm}[t]
  \DontPrintSemicolon
  \SetKwInOut{Input}{Input}\SetKwInOut{Output}{Output}
  \Input{$h,T,D_T(h),\bold{B},k$\textnormal{, learning rate }$\mu$\textnormal{, threshold }$\epsilon$}
  \Output{$\omega$\textnormal{, top-}$k$ \textnormal{most prominent entities.}}

\textnormal{ }\\
Initialize: $\omega\coloneqq\omega^{(0)}$ \\
Calculate $\bold{f}_m,\bold{f}_c,\bold{f}_t, \bold{f}_{\omega}\coloneqq\bold{f}_{\omega^{(0)}}$ using Eqs. \ref{eq:of}, \ref{eq:m}, \ref{eq:c}, \ref{eq:t} \\
\While{true}
{
    $\hat{\bold{f}}_{\omega} \coloneqq $ \textnormal{normalize} $\bold{f}_{\omega}$ \\
    Set $\bold{s_h}\coloneqq\hat{\bold{f}}_{\omega},$ \textnormal{calculate} $\bold{r_h}$  \textnormal{using Eq.~\ref{eq:rw}} \\
    Sort $\bold{r_h},$ \textnormal{get the top-}$k$ \textnormal{entities} $E(h,k)$   \\
    \If{$\sum_{e\in E(h,k)}{L(f(e,h),r(e,h))} < \epsilon$}
    {
        Stop   
    }
    $\omega \coloneqq \omega - \mu\sum_{e\in E(h,k)}{\nabla L(f(e,h),r(e,h))}$
}

\textbf{return} $\omega,E(h,k)$\;

\caption{Entity Influence-Prominence Learning}
 \label{alg:pil}
\end{algorithm}


The main steps are depicted in Algorithm~\ref{alg:pil}. We start with
an initial guess for $\omega$, and compute the similarities for the
candidate entities. Here $\bold{f}_m$, $\bold{f}_c$, $\bold{f}_t$, and
$\bold{f}_\omega$ represent the similarity score vectors. We use
matrix multiplication to calculate the similarities efficiently. In
each iteration, we first normalize $\bold{f}_\omega$ such that the
entity scores sum up to $1$. A random walk is performed to calculate
the influence score $\bold{r_h}$. Then we update $\omega$ using a
batch gradient descent method on the top-$k$ influencer entities. To
derive the gradient of the loss function $L$, we first remark that our
random walk Equation~\ref{eq:rw} is similar to context-sensitive
PageRank~\cite{Haveliwala2002}. Using the linearity
property~\cite{fogaras2005towards}, we can express $r(e,h)$ as the
linear function of influence scores obtained by initializing with the
individual similarities $f_m, f_c$, and $f_t$ instead of
$f_\omega$. The derivative thus can be written as:
\begin{flalign*}
&\nabla L(f(e,h),r(e,h)) = \alpha(r_m(e,h)-f_m(e,h))+&\\
&\qquad\beta(r_c(e,h)-f_c(e,h))+\gamma(r_t(e,h)-f_t(e,h))
\end{flalign*}
%
where $r_m(e,h),r_c(e,h),r_t(e,h)$ are the components of the three vector solutions of Equation~\ref{eq:rw}, each having $\bold{s_h}$ replaced by $\bold{f}_m$, $\bold{f}_c$, $\bold{f}_t$ respectively.

Since both $\bold{B}$ and $\hat{\bold{f}_{\omega}}$ are normalized
such that their column sums are equal to $1$, Equation~\ref{eq:rw} is
convergent~\cite{Haveliwala2002}. Also, as discussed above,
$\bold{r_h}$ is a linear combination of factors that are independent
of $\omega$, hence $L$ is a convex function, and the batch gradient
descent is also guaranteed to converge. In practice, we can utilize
several indexing techniques to significantly speed up the similarity
and influence scores calculation.


\section{Experiments and Results}

\subsection{Setup}\label{sec:prep}

\begin{table}
  \centering
  \begin{tabular}{@{}l r @{}}
    \toprule
    Total Tweets                  &  500,551,041 \\
    Trending Hashtags             &  2,444 \\
    Test Hashtags                 &  30 \\
    Test Tweets                   &  352,394 \\
    Distinct Mentions             &  145,941 \\
    Test (Entity, Hashtag) pairs  &  6,965 \\
    Candidates per Hashtag (avg.) &  50 \\
    Extended Candidates (avg.)    & 182 \\
    \toprule
  \end{tabular}
  \captionsetup{justification=centering}
  \caption{Statistics of the dataset.}\label{table:fa}
\end{table}


\begin{table*}[!tb]
  \centering
  \begin{tabular}{@{}lcccccccc@{}}
    \toprule
         & Tagme & Wikiminer & Meij          & Kauri & M     & C     & T     & IPL\\
    \midrule
    P@5  & 0.284 & 0.253     & 0.500          & 0.305 & 0.453 & 0.263 & 0.474 & \textbf{0.642} \\
    P@15 & 0.253 & 0.147     & \textbf{0.670} & 0.319 & 0.312 & 0.245 & 0.378 & 0.495 \\
    MAP  & 0.148 & 0.096     & 0.375         & 0.162 & 0.211 & 0.140 & 0.291 & \textbf{0.439} \\
    \bottomrule
  \end{tabular}
  \captionsetup{justification=centering,margin=2cm}
  \caption{Experimental results on the sampled trending hashtags.}  \label{tbl:result}
\end{table*}


\paragraph{Dataset} There is no standard benchmark for our problem,
since available datasets on microblog annotation (such as the
Microposts challenge \cite{microposts2014_neel_cano.ea:2014}) do not
have global statistics, so we cannot identify the trending hashtags. Therefore, we created our own dataset. We used the Twitter
API to collect from the public stream a sample of $500,551,041$ tweets
from January to April 2014. We removed hashtags that were adopted by
less than $500$ users, having no letters, or having characters
repeated more than $4$ times (e.g., `\#oooommgg'). We identified
trending hashtags by computing the daily time series of hashtag tweet
counts, and removing those of which the time series' variance score is
less than $900$. To identify the hashtag burst time period $T$, we
compute the \emph{outlier fraction} \cite{lehmann2012dynamical} for
each hashtag $h$ and day $t$:
$p_t(h)=\frac{|n_t-n_b|}{\max{(n_b,n_{\min})}}$,
where $n_t$ is the number of tweets containing $h$, $n_b$ is the
median value of $n_t$ over all points in a $2$-month time window
centered on $t$, and $n_{\min}=10$ is the threshold to filter low
activity hashtags. The hashtag is skipped if its highest outlier
fraction score is less than $15$. Finally, we define the \emph{burst
  time period} of a trending hashtag as the time window of size $w$,
centered at day $t_0$ with the highest $p_{t_{0}}(h)$.

%

For the Wikipedia datasets we process the dump from 3rd May 2014, so
as to cover all events in the Twitter dataset. We have developed
Hedera \cite{traniswc14}, a scalable tool for processing the Wikipedia
revision history dataset based on Map-Reduce paradigm. In
addition, we download the Wikipedia page view dataset that stores how
many times a Wikipedia article was requested on an hourly level. We
process the dataset for the four months of our study and use Hedera to
accumulate all view counts of redirects to the actual articles.

\paragraph{Sampling} From the trending hashtags, we sample $30$
distinct hashtags for evaluation. Since our study focuses on trending
hashtags that are mapable to entities in Wikipedia, the sampling must
cover a sufficient number of ``popular'' topics that are seen in
Wikipedia, and at the same time cover rare topics in the long tail. To do
this, we apply several heuristics in the sampling. First, we only
consider hashtags where the lexicon-based linking
(Section~\ref{subsec:el}) results in at least 20 different
entities. Second, we randomly choose hashtags to cover different types
of topics (long-running events, breaking events, endogenous
hashtags). Instead of inspecting all hashtags in our corpus, we follow
\newcite{lehmann2012dynamical} and calculate the fraction of tweets
published before, during and after the peak. The hashtags are then
clustered in this 3-dimensional vector space. Each cluster suggests a
group of hashtags with a distinct
semantics~\cite{lehmann2012dynamical}. We then pick up hashtags
randomly from each cluster, resulting in 200 hashtags in total. From
this rough sample, three inspectors carefully checked the tweets and
chose 30 hashtags where the meanings and hashtag types were certain to
the knowledge of the inspectors.
\vspace{-0.5em}
\paragraph{Parameter Settings} We initialize the similarity weights to
$\frac{1}{3}$, the damping factor to $\tau=0.85$, and the weight for
the language model to $\lambda=0.9$. The learning rate $\mu$ is
empirically fixed to $\mu=0.003$.

\paragraph{Baseline} We compare IPL with other entity annotation
methods. Our first group of baselines includes entity linking systems
in domains of general text, Wikiminer~\cite{milne2008learning}, and
short text, Tagme~\cite{Ferragina:2012:FAA:2122259.2122384}. For each
method, we use the default parameter settings, apply them for the
individual tweets, and take the average of the annotation confidence
scores as the prominence ranking function. The second group of
baselines includes systems specifically designed for microblogs. For
the content-based methods, we compare against~\newcite{meij2012adding},
which uses a supervised method to rank entities with respect to
tweets. We train the model using the same training data as in the
original paper. For the graph-based method, we compare against
KAURI~\cite{shen2013linking}, a method which uses user interest
propagation to optimize the entity linking scores. To tune the
parameters, we pick up four hashtags from different clusters, randomly
sample 50 tweets for each, and manually annotate the tweets. For all baselines, we obtained the implementation from the authors. The exception is Meij method, where we implemented ourselves, but we clarified with the authors via emails on several settings. In addition, we also
compare three variants of our method, using only local functions for
entity ranking (referred to as $M$, $C$, and $T$ for \emph{mention},
\emph{context}, and \emph{time}, respectively).
\vspace{-0.5em}
\paragraph{Evaluation} In total, there are $6,965$ entity-hashtag
pairs returned by all systems. We employ five volunteers to evaluate
the pairs in the range from $0$ to $2$, where $0$ means the entity is
noisy or obviously unrelated, $2$ means the entity is strongly tied to
the topic of the hashtag, and $1$ means that although the entity and
hashtag might share some common contexts, they are not involved in a
direct relationship (for instance, the entity is a too general concept
such as \wpent{Ice hockey}, as in the case illustrated in
Figure~\ref{fig:infgraph}). The annotators were advised to use search
engines, the Twitter search box or Wikipedia archives whenever
applicable to get more background on the stories. Inter-annotator
agreement under Fleiss score is $0.625$.

\subsection{Results and Discussion}
Table~\ref{tbl:result} shows the performance comparison of the methods
using the standard metrics for a ranking system (precision at 5 and 15
and MAP at 15). In general, all baselines perform worse than reported
in the literature, confirming the higher complexity of the hashtag
annotation task as compared to traditional tasks. Interestingly
enough, using our local similarities already produces better results
than Tagme and Wikiminer. The local model $f_m$ significantly
outperforms both the baselines in all metrics. Combining the
similarities improves the performance even more
significantly.\footnote{All significance tests are done against both
  Tagme and Wikiminer, with a $p$-value $<0.01$.} Compared to the
baselines, IPL improves the performance by 17-28\%.  The time
similarity achieves the highest result compared to other content-based
mention and context similarities. This supports our assumption that
lexical matching is not always the best strategy to link entities in
tweets. The time series-based metric incurs lower cost than others,
yet it produces a considerably good performance. Context similarity
based on Wikipedia edits does not yield much improvement. This can be
explained in two ways. First, information in Wikipedia is largely
biased to popular entities, it fails to capture many entities in the
long tail.
Second, language models are dependent on direct word representations,
which are different between Twitter and Wikipedia.  This is another
advantage of non-content measures such as $f_t$.

For the second group of baselines (Kauri and Meij), we also observe
the reduction in precision, especially for Kauri. This is because the
method relies on the coherence of user interests within a group of
tweets to be able to perform well, which does not hold in the context
of hashtags. One astonishing result is that Meij performs better than
IPL in terms of P@15. However, it performs worse in terms of MAP and
P@5, suggesting that most of the correctly identified entities are
ranked lower in the list. This is reasonable, as Meij attempts to
optimize (with human supervision effort) the semantic agreement
between entities and information found in the tweets, instead of
ranking their prominence as in our work. To investigate this case
further, we re-examined the hashtags and divided them by their
semantics, as to whether the hashtags are spurious trends of memes
inside social media (\emph{endogenous}, e.g., ``\#stopasian2014''), 
or whether they reflect external events (\emph{exogenous}, e.g., ``\#mh370''). 
%
\begin{figure}[!tb]
  \centering
  \includegraphics[width=\columnwidth]{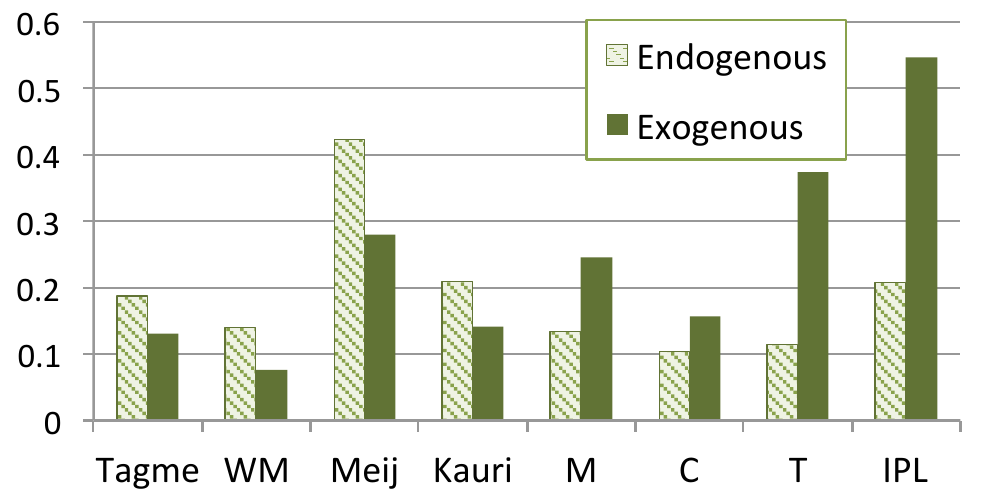}
  \caption{Performance of the methods for different types of trending hashtags.}
  \label{fig:endoExo}
\end{figure}
The performance of the methods in terms of MAP scores is shown in
Figure~\ref{fig:endoExo}. It can be clearly seen that entity linking
methods perform well in the endogenous group, but then deteriorate in
the exogenous group. The explanation is that for endogenous hashtags,
the topical consonance between tweets is very low, 
thus most of the assessments become just verifying general concepts (such as locations)
In this case, topical annotation is trumped by conceptual annotation. However,
whenever the hashtag evolves into a meaningful topic, a deeper
annotation method will produce a significant improvement, as seen in
Figure~\ref{fig:endoExo}.

\begin{figure}[!tb]
  \centering
  \includegraphics[width=\linewidth]{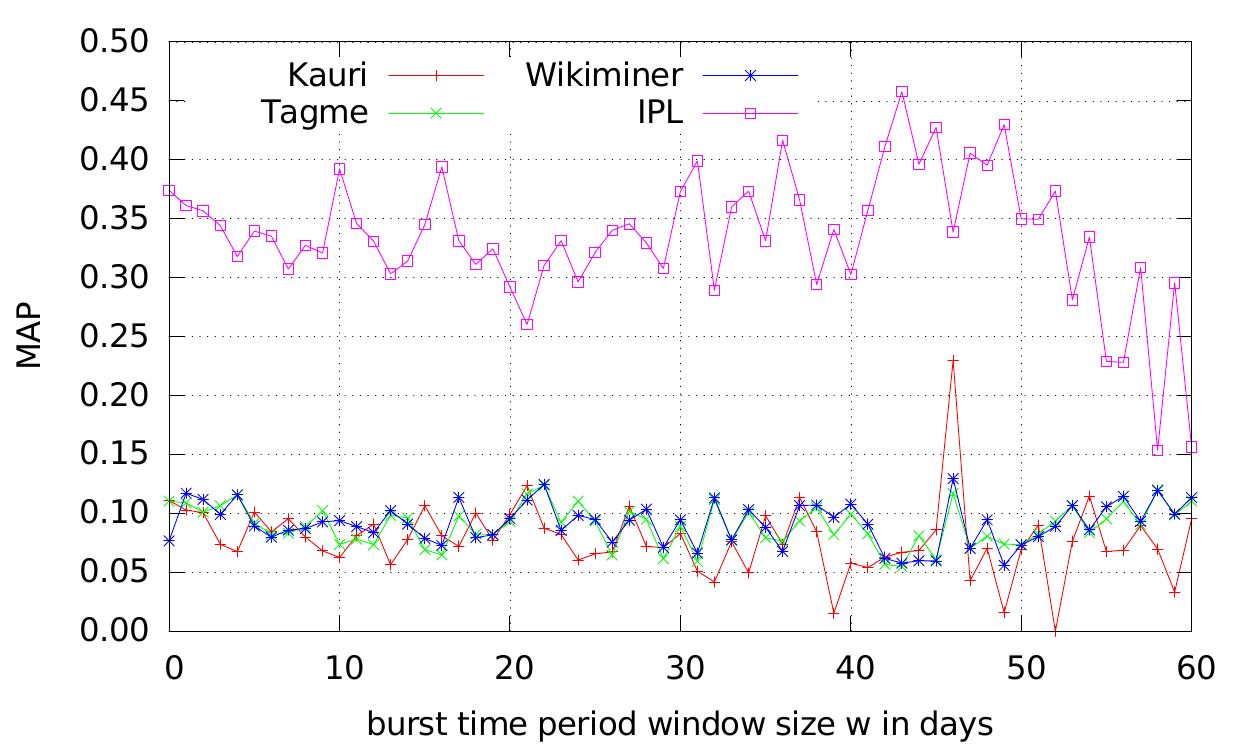}
  \caption{IPL compared to other baselines on different sizes of the burst time window $T$.}
  \label{fig:window}
\end{figure}

Finally, we study the impact of the burst time period on the
annotation quality. For this, we expand the window size $w$
(cf. Section~\ref{sec:prep}) and examine how different methods
perform. The result is depicted in Figure~\ref{fig:window}. It is
obvious that within the window of $2$ months (where the hashtag time
series is constructed and a trending time is identified), our method
is stable and always outperforms the baselines by a large margin. Even
when the trending hashtag has been saturated, hence introduced more
noise, our method is still able to identify the prominent entities
with high quality.


\section{Conclusion and Future Work}\label{sec:concl-future-work}

In this work, we address the new problem of topically annotating a
trending hashtag using Wikipedia entities, which has many important
applications in social media analysis. We study Wikipedia temporal
resources and find that using efficient time series-based measures can
complement content-based methods well in the domain of Twitter. We
propose use similarity measures to model both the local
mention-based, as well as the global context- and time-based
prominence of entities. We propose a novel strategy of topical
annotation of texts using and influence maximization approach and
design an efficient learning algorithm to automatically unify the
similarities without the need of human involvement. The experiments
show that our method outperforms significantly the established
baselines.

As future work, we aim to improve the efficiency of our entire
workflow, such that the annotation can become an end-to-end
service. We also aim to improve the context similarity between
entities and the topic, for example by using a deeper distributional
semantics-based method, instead of language models as in our current
work. In addition, we plan to extend the annotation framework to other
types of trending topics, by including the type of out-of-knowledge
entities. Finally, we are investigating how to apply more advanced
influence maximization methods. We believe that influence maximization has a great
potential in NLP research, beyond the scope of annotation for microblogging topics.


\section*{Acknowledgments}
This work was funded by the European Commission in the FP7 project
ForgetIT (600826) and the ERC advanced grant ALEXANDRIA (339233), and
by the German Federal Ministry of Education and Research for the
project ``Gute Arbeit'' (01UG1249C). We thank the reviewers for the
fruitful discussion 
and Claudia Niederee from L3S for suggestions on improving
Section~\ref{sec:im}.


\bibliographystyle{acl}

\begin{thebibliography}{}

\bibitem[\protect\citename{Bansal \bgroup et al.\egroup }2015]{BansalBV15}
P.~Bansal, R.~Bansal, and V.~Varma.
\newblock 2015.
\newblock Towards deep semantic analysis of hashtags.
\newblock In {\em {ECIR}}, pages 453--464.

\bibitem[\protect\citename{Basave \bgroup et al.\egroup
  }2014]{microposts2014_neel_cano.ea:2014}
A.~E.~Cano Basave, G.~Rizzo, A.~Varga, M.~Rowe, M.~Stankovic, and A.~Dadzie.
\newblock 2014.
\newblock Making sense of microposts ({\#microposts2014}) named entity
  extraction \& linking challenge.
\newblock In {\em 4th Workshop on Making Sense of Microposts}.

\bibitem[\protect\citename{Cassidy \bgroup et al.\egroup
  }2012]{cassidy2012analysis}
T.~Cassidy, H.~Ji, L.-A. Ratinov, A.~Zubiaga, and H.~Huang.
\newblock 2012.
\newblock Analysis and enhancement of wikification for microblogs with context
  expansion.
\newblock In {\em {COLING}}, pages 441--456.

\bibitem[\protect\citename{Ciglan and N{\o}rv{\aa}g}2010]{ciglan2010wikipop}
M.~Ciglan and K.~N{\o}rv{\aa}g.
\newblock 2010.
\newblock {WikiPop}: personalized event detection system based on {Wikipedia}
  page view statistics.
\newblock In {\em {CIKM}}, pages 1931--1932.

\bibitem[\protect\citename{Fang and Chang}2014]{fang2014entity}
Y.~Fang and M.-W. Chang.
\newblock 2014.
\newblock Entity linking on microblogs with spatial and temporal signals.
\newblock {\em {Trans. of the Assoc. for Comp. Linguistics}}, 2:259--272.

\bibitem[\protect\citename{Ferragina and
  Scaiella}2012]{Ferragina:2012:FAA:2122259.2122384}
P.~Ferragina and U.~Scaiella.
\newblock 2012.
\newblock Fast and accurate annotation of short texts with {Wikipedia} pages.
\newblock {\em IEEE Softw.}, 29(1):70--75.

\bibitem[\protect\citename{Fogaras \bgroup et al.\egroup
  }2005]{fogaras2005towards}
D.~Fogaras, B.~R{\'a}cz, K.~Csalog{\'a}ny, and T.~Sarl{\'o}s.
\newblock 2005.
\newblock Towards scaling fully personalized {PageRank}: Algorithms, lower
  bounds, and experiments.
\newblock {\em Internet Mathematics}, 2(3):333--358.

\bibitem[\protect\citename{Guo \bgroup et al.\egroup }2013]{guo2013link}
S.~Guo, M.-W. Chang, and E.~K{\i}c{\i}man.
\newblock 2013.
\newblock To link or not to link? {A} study on end-to-end tweet entity linking.
\newblock In {\em {NAACL-HLT}}, pages 1020--1030.

\bibitem[\protect\citename{Haveliwala}2002]{Haveliwala2002}
T.~H. Haveliwala.
\newblock 2002.
\newblock Topic-sensitive {PageRank}.
\newblock In {\em {WWW}}, pages 517--526.

\bibitem[\protect\citename{Keegan \bgroup et al.\egroup
  }2011]{Keegan:2011:HOW:2038558.2038577}
Brian Keegan, Darren Gergle, and Noshir Contractor.
\newblock 2011.
\newblock Hot off the wiki: Dynamics, practices, and structures in wikipedia's
  coverage of the t\-{o}hoku catastrophes.
\newblock In {\em {WikiSym}}, pages 105--113.

\bibitem[\protect\citename{Kempe \bgroup et al.\egroup
  }2003]{kempe2003maximizing}
D.~Kempe, J.~Kleinberg, and {\'E}.~Tardos.
\newblock 2003.
\newblock Maximizing the spread of influence through a social network.
\newblock In {\em {KDD}}, pages 137--146.

\bibitem[\protect\citename{Lappas \bgroup et al.\egroup }2009]{Lappas:2009}
T.~Lappas, B.~Arai, M.~Platakis, D.~Kotsakos, and D.~Gunopulos.
\newblock 2009.
\newblock On burstiness-aware search for document sequences.
\newblock In {\em {KDD}}, pages 477--486.

\bibitem[\protect\citename{Lehmann \bgroup et al.\egroup
  }2012]{lehmann2012dynamical}
J.~Lehmann, B.~Gon{\c{c}}alves, J.~J. Ramasco, and C.~Cattuto.
\newblock 2012.
\newblock Dynamical classes of collective attention in {Twitter}.
\newblock In {\em {WWW}}, pages 251--260.

\bibitem[\protect\citename{Liu \bgroup et al.\egroup }2013]{liu2013entity}
X.~Liu, Y.~Li, H.~Wu, M.~Zhou, F.~Wei, and Y.~Lu.
\newblock 2013.
\newblock Entity linking for tweets.
\newblock In {\em {ACL}}, pages 1304--1311.

\bibitem[\protect\citename{Liu \bgroup et al.\egroup }2014]{liu2014influence}
Q.~Liu, B.~Xiang, E.~Chen, H.~Xiong, F.~Tang, and J.~X. Yu.
\newblock 2014.
\newblock Influence maximization over large-scale social networks: {A} bounded
  linear approach.
\newblock In {\em {CIKM}}, pages 171--180.

\bibitem[\protect\citename{Meij \bgroup et al.\egroup }2012]{meij2012adding}
E.~Meij, W.~Weerkamp, and M.~de~Rijke.
\newblock 2012.
\newblock Adding semantics to microblog posts.
\newblock In {\em {WSDM}}, pages 563--572.

\bibitem[\protect\citename{Milne and Witten}2008]{milne2008learning}
D.~Milne and I.~H. Witten.
\newblock 2008.
\newblock Learning to link with {Wikipedia}.
\newblock In {\em {CIKM}}, pages 509--518.

\bibitem[\protect\citename{Naaman \bgroup et al.\egroup }2011]{naaman2011hip}
M.~Naaman, H.~Becker, and L.~Gravano.
\newblock 2011.
\newblock Hip and trendy: Characterizing emerging trends on {Twitter}.
\newblock {\em {JASIST}}, 62(5):902--918.

\bibitem[\protect\citename{Osborne \bgroup et al.\egroup
  }2012]{osborne2012bieber}
M.~Osborne, S.~Petrovic, R.~McCreadie, C.~Macdonald, and I.~Ounis.
\newblock 2012.
\newblock Bieber no more: {First} story detection using {Twitter} and
  {Wikipedia}.
\newblock In {\em Workshop on Time-aware Information Access}.

\bibitem[\protect\citename{Owoputi \bgroup et al.\egroup
  }2013]{owoputi2013improved}
O.~Owoputi, B.~O'Connor, C.~Dyer, K.~Gimpel, N.~Schneider, and N.~A. Smith.
\newblock 2013.
\newblock Improved part-of-speech tagging for online conversational text with
  word clusters.
\newblock In {\em {NAACL-HLT}}, pages 380--390.

\bibitem[\protect\citename{Radinsky \bgroup et al.\egroup }2011]{Radinsky2011}
K.~Radinsky, E.~Agichtein, E.~Gabrilovich, and S.~Markovitch.
\newblock 2011.
\newblock A word at a time: {Computing} word relatedness using temporal
  semantic analysis.
\newblock In {\em {WWW}}, pages 337--346.

\bibitem[\protect\citename{Ratinov \bgroup et al.\egroup
  }2011]{Ratinov:2011:LGA:2002472}
L.~Ratinov, D.~Roth, D.~Downey, and M.~Anderson.
\newblock 2011.
\newblock {Local and Global Algorithms for Disambiguation to {Wikipedia}}.
\newblock In {\em {ACL}}, pages 1375--1384.

\bibitem[\protect\citename{Ritter \bgroup et al.\egroup }2011]{ritter2011named}
Alan Ritter, Sam Clark, Oren Etzioni, et~al.
\newblock 2011.
\newblock Named entity recognition in tweets: an experimental study.
\newblock In {\em {EMNLP}}, pages 1524--1534.

\bibitem[\protect\citename{Shen \bgroup et al.\egroup }2013]{shen2013linking}
W.~Shen, J.~Wang, P.~Luo, and M.~Wang.
\newblock 2013.
\newblock Linking named entities in tweets with knowledge base via user
  interest modeling.
\newblock In {\em {WSDM}}, pages 68--76.

\bibitem[\protect\citename{Tolomei \bgroup et al.\egroup }2013]{Tolomei2013TAB}
G.~Tolomei, S.~Orlando, D.~Ceccarelli, and C.~Lucchese.
\newblock 2013.
\newblock Twitter anticipates bursts of requests for {Wikipedia} articles.
\newblock In {\em Workshop on Data-driven User Behavioral Modelling and Mining
  from Social Media}, pages 5--8.

\bibitem[\protect\citename{Tran and Nguyen}2014]{traniswc14}
T.~Tran and T.~Ngoc Nguyen.
\newblock 2014.
\newblock Hedera: {Scalable} indexing, exploring entities in {Wikipedia}
  revision history.
\newblock In {\em {ISWC}}, pages 297--300.

\bibitem[\protect\citename{Tran \bgroup et al.\egroup }2014]{Tran2014Wikitweet}
T.~Tran, M.~Georgescu, X.~Zhu, and N.~Kanhabua.
\newblock 2014.
\newblock Analysing the duration of trending topics in {Twitter} using
  {Wikipedia}.
\newblock In {\em Conf. on Web Science}, pages 251--252.

\bibitem[\protect\citename{Tsur and Rappoport}2012]{Tsur:2012}
O.~Tsur and A.~Rappoport.
\newblock 2012.
\newblock What's in a hashtag?: Content based prediction of the spread of ideas
  in microblogging communities.
\newblock In {\em {WSDM}}, pages 643--652.

\bibitem[\protect\citename{Wang \bgroup et al.\egroup }2011]{wang2011web}
K.~Wang, C.~Thrasher, and B.-J.~P. Hsu.
\newblock 2011.
\newblock Web scale {NLP}: a case study on {URL} word breaking.
\newblock In {\em {WWW}}, pages 357--366.

\bibitem[\protect\citename{Yang and Leskovec}2011]{yang2011patterns}
J.~Yang and J.~Leskovec.
\newblock 2011.
\newblock Patterns of temporal variation in online media.
\newblock In {\em {WSDM}}, pages 177--186.

\end{thebibliography}

\end{document}